\documentclass[graybox]{svmult}

\usepackage{mathptmx}       
\usepackage{helvet}         
\usepackage{courier}        
\usepackage{type1cm}        
\usepackage{makeidx}         
\usepackage{graphicx}        
\usepackage{multicol}        
\usepackage[bottom]{footmisc}
\usepackage{amssymb}

\makeindex             

\begin{document}

\title*{Cloud and Star Formation in Spiral Arms}
\author{Clare Dobbs and Alex Pettitt}
\institute{Clare Dobbs \at School of Physics \& Astronomy, University of Exeter, EX4 4QL, UK, \email{dobbs@astro.ex.ac.uk}
\and Alex Pettitt \at School of Physics \& Astronomy, University of Exeter, EX4 4QL, UK, \email{alex@astro.ex.ac.uk}}
%
%
\maketitle

\abstract{We present the results from simulations of GMC formation in spiral galaxies. First we discuss cloud formation by cloud--cloud collisions, and gravitational instabilities, arguing that the former is prevalent at lower galactic surface densities and the latter at higher. Cloud masses are also limited by stellar feedback, which can be effective before clouds reach their maximum mass. We show other properties of clouds in simulations with different levels of feedback. With a moderate level of feedback, properties such as cloud rotations and virial parameters agree with observations. Without feedback, an unrealistic population of overly bound clouds develops. Spiral arms are not found to trigger star formation, they merely gather gas into more massive GMCs. We discuss in more detail interactions of clouds in the ISM, and argue that these are more complex than early ideas of cloud--cloud collisions. Finally we show ongoing work to determine whether the Milky Way is a flocculent or grand design spiral.}
\section{Introduction}
\label{sec:1}
Theoretically, there are 3 main mechanisms for the formation of Giant Molecular Clouds (GMCs) in galaxies; gravitational instabilities, cloud-cloud collisions and Parker instabilities. In the past decade or so, numerical simulations have explored these scenarios. Here I focus on cloud-cloud collisions and gravitational instabilities, the regimes where these mechanisms dominate, and the properties of clouds that are predicted. I will also argue that although historically cloud-cloud collisions have been proposed as an important mechanism, in the ISM, interactions of clouds and diffuse gas are more complex than this simple description.

\section{Simulations of cloud-cloud collisions, gravitational instabilities and stellar feedback in grand design spirals}
\label{sec:2}
\begin{figure}[b]
\sidecaption[t]
\includegraphics[scale=.26]{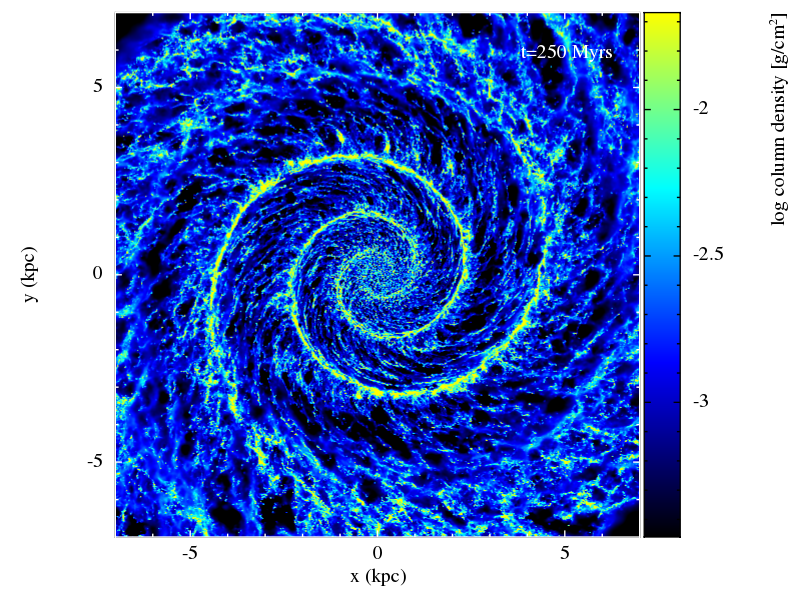}
\caption{Example simulation of a galaxy with an $m=2$ imposed spiral potential. The simulation includes self gravity, cooling and heating of the ISM and stellar feedback (producing a multiphase ISM), but not magnetic fields. Dense regions shown by yellow in the colour table correspond to molecular clouds.}
\label{fig:1}       
\end{figure}

Cloud-cloud collisions were originally investigated in relatively simplistic calculations which follow a population of typically spherical clouds, employing  a statistical description for the the formation of small clouds, the coalescence of clouds, and collapse of clouds above a given size (Field \& Saslaw 1965; Scoville \& Hersch 1979; Kwan \& Valdes 1983, 1987; Tomisaka 1984, 1986). Most assume clouds completely coalesce, although Roberts \& Stewart 1987 allow dissipative collisions, which results in clouds clustering together at points along the spiral arms. In various papers (Dobbs et al. 2006; Dobbs 2008; Dobbs et al. 2008) I demonstrated that GMCs can build up from smaller scale structure in hydrodynamic simulations. Instead of the Jeans length and mass, the sizes and masses of clouds formed in this way is dependent on the strength of the spiral shock. In particular the epicyclic radius which determines the amount of mass that can be gathered together into a single GMC. This mechanism is most evident in discs where the gas is not strongly gravitationally unstable, found to be when $\Sigma_{gas}\lesssim10$ M$_{\odot}$pc$^{-2}$. At higher surface densities, gravitational instabilities start to dominate in the disc, although self gravity also increases collisions between clouds and GMCs. We show an example from a simulation with $\Sigma_{gas}=8$ M$_{\odot}$pc$^{-2}$, where both cloud--cloud collisions and self gravity are significant, in Figure~\ref{fig:1} (from Dobbs \& Pringle 2013).

One criticism of the cloud-cloud collision model for forming massive GMCs is the long timescale thought to be required (Blitz \& Shu 1980), of order 100s Myrs. However in a galaxy with spiral arms, the cloud number density is significantly increased in the spiral arms (Casoli \& Combes 1981; Dobbs 2008). The time for the growth of a GMC can be estimated from the mean free path divided by the cloud--cloud velocity dispersion, giving 
\begin{equation}
t=\frac{1}{\pi r_{cl}^2 n_{cl} \sigma}
\end{equation}
where $r_{cl}$ is an average cloud radius, $n_{cl}$ is the cloud number density and $\sigma$ is the cloud-cloud velocity dispersion. With values of $r_{cl}=30$ pc, $\sigma=4$ km/s and an average $n_{cl}=9 \times 10^{-8}$ clouds per pc$^3$ (based on the numerical simulation of Dobbs \& Pringle 2013), this gives a timescale of $\sim$ 900 Myr. However in the spiral arms, $n_{cl}$ can be 20 or more times higher giving timescales of 10 or a few 10's Mys. Work by Fujimoto et al. 2014 also finds similar timescales. 

A limitation of this previous work though is that in the absence of stellar feedback, the star formation rate is orders of magnitude too high (see Bonnell et al. 2011; van Loo et al. 2011;  Dobbs et al. 2011; Kim et al. 2013), whilst stellar feedback may also disperse clouds (Elmegreen 1994). We included a simple treatment of stellar feedback by inserting kinetic and thermal energy according to a Sedov solution when star formation is assumed to occur. The amount of energy inserted is given by 
\begin{equation}
E=\frac{\epsilon M(H_2) \times 10^{51} ergs}{160 M_{\odot}}
\end{equation}
where M(H$_2$) is the mass of molecular hydrogen within roughly a smoothing length, the 160 M$_{\odot}$ assumes that one massive star forms per 160 M$_{\odot}$ of gas, and the $10^{51}$ ergs is assumed to be the energy released from one massive star. $\epsilon$ is the star formation efficiency at the resolution of the simulation (few $10^4$ M$_{\odot}$ for particle masses of 312.5 M$_{\odot}$), our fiducial value is 5 \%. Energy is inserted instantaneously and is presumed to be associated with stellar winds and supernovae. In Dobbs et al. 2011, we indeed showed that the cloud mass spectra are shifted to lower masses depending on the level of stellar feedback. We also find that the star formation rate is in much better agreement with observations if stellar feedback is included (Dobbs et al. 2011). Generally the properties of clouds vary according to the level of feedback as described in the next section.

\subsection{Properties of GMCs in simulations}
We considered some properties of GMCs in the absence of stellar feedback in Dobbs 2008. In particular we showed that in a low surface density case, where self gravity has less effect, clouds exhibit retrograde as well as prograde rotation. Although we did not show the comparison for the high surface density case, there were comparatively more prograde clouds. We have since more extensively considered cloud properties in Dobbs et al. 2011, with stellar feedback. Here we showed a transition between gravitationally dominated clouds in simulations with minimal feedback ($\epsilon=0.01$) compared to clouds in simulations where there is roughly a balance between stellar feedback, self gravity and the spiral shock (for our simulations $\epsilon=0.05$). The surface density in these calculations was 8 M$_{\odot}$ pc$^{-2}$.
\begin{figure}[t]
\includegraphics[scale=.55]{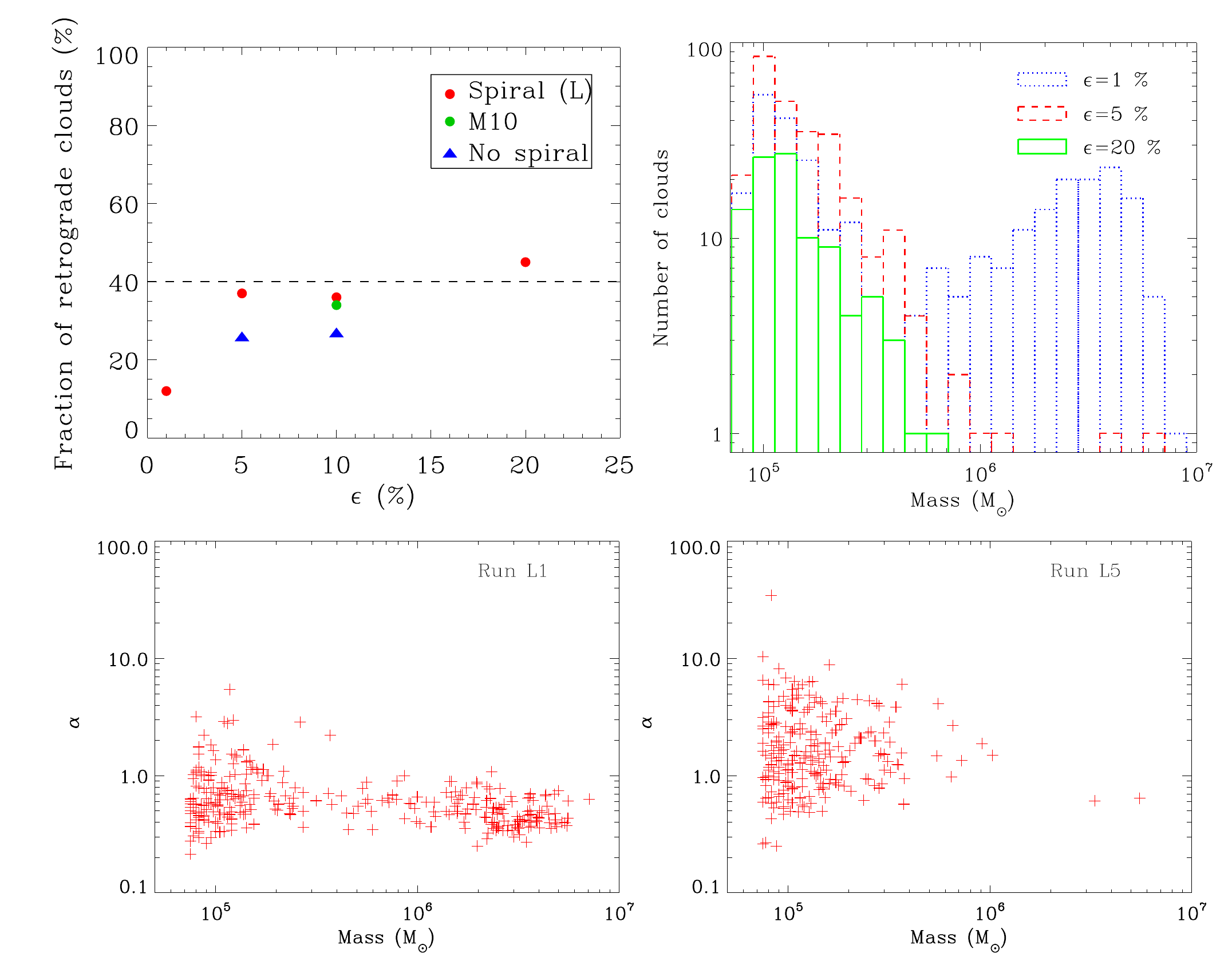}
\caption{Properties of clouds are shown from simulations in Dobbs et al. 2011. The top left panel shows the fraction of retrograde clouds versus the level of stellar feedback. The top right panel shows mass spectra for different levels of feedback. The lower panels show the virial parameter when $\epsilon=0.01$ (left, minimal feedback) and 0.05 (right, moderate feedback).}
\label{fig:2}       
\end{figure}

We show 3 cloud properties in Figure~\ref{fig:2}; the cloud mass spectra, the cloud virial parameters, and the fraction of retrograde clouds, for simulations with varying levels of feedback. Figure~\ref{fig:2} shows a noticeable difference between clouds with and without significant feedback. Without stellar feedback, the clouds eventually become unable to disperse by shear and generic turbulence alone, and become gravitationally dominated. Gas eventually either falls into massive GMCs or lies in the warm component but there is no interaction between the components, and minimal interactions between clouds. This results in a mass spectrum centred on $10^6-10^7$ M$_{\odot}$ clouds . By contrast, with feedback, the mass spectra display a typical $\sim$M$^{-2}$ power law, clearly in much better agreement with observations, the level of feedback determining the maximum mass. 

Figure~\ref{fig:2} also shows cloud rotations for different levels of stellar feedback. For our standard spiral galaxy models (red points), the fraction of retrograde rotating clouds is low with minimal feedback, but roughly constant when $\epsilon \geq 0.05$. Again this is because in the minimal feedback case, the gas is confined in clouds which are not interacting with each other. Finally we show the virial parameters of the clouds in Figure~\ref{fig:2} in simulations with $\epsilon=0.01$ and $0.05$. The distribution of $\alpha$ is less than 1 when $\epsilon=0.01$ indicating that the clouds are strongly gravitationally bound. With stellar feedback, the velocity dispersions of the gas are generally higher, meaning gas is less bound, and gas is also ejected from clouds before it can become too gravitationally bound.

\subsection{The nature of `cloud-cloud collisions'}
The term `cloud-cloud collisions' is commonly used to describe interactions of clouds in a galaxy, to form more massive GMCs or induce star formation (Tan 2000). However in reality the ISM is a continuum, rather than divided separately into cold clouds surrounded by a warm diffuse medium. According to simulations, a substantial fraction of gas is in an intermediate regime between cold and warm ISM (Gazol et al. 2001, Dobbs et al. 2011) whilst transitions of gas between cold, intermediate, warm, and molecular phases may be relatively frequent. Furthermore interactions between clouds may not be particularly violent, or induce much change in momentum. Interactions can take the form of grazing collisions, where there is little interaction or mass transfer between the clouds, or full mergers, where the two clouds collide and all the mass from both clouds end up in one resultant cloud.
\begin{figure}[t]
\includegraphics[scale=.65]{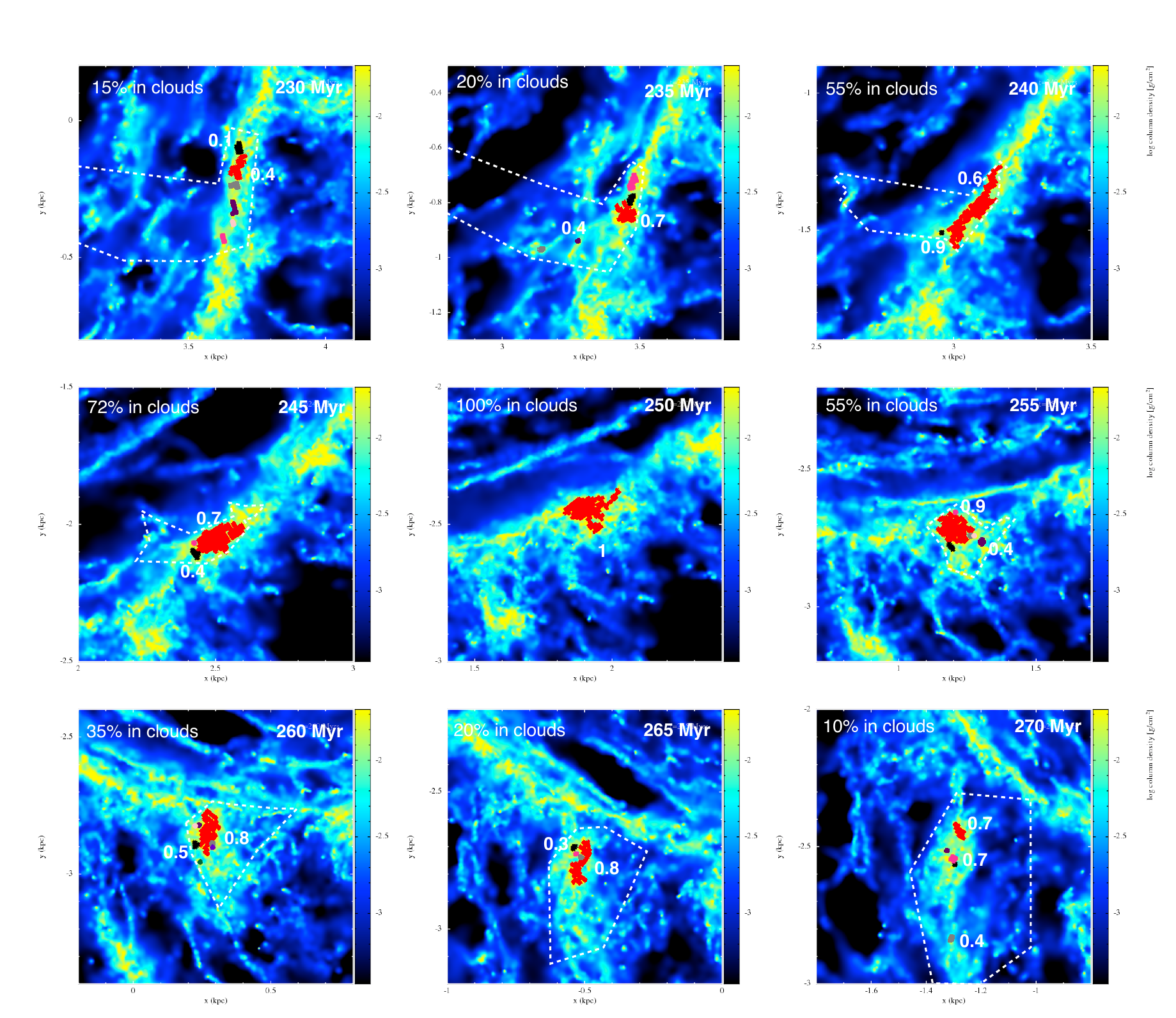}
\caption{The evolution of a $2\times10^6$ M$_{\odot}$ GMC (coloured red) is shown over a 40 Myr time period. Other clouds which coalesce with this GMC, or are the result of dispersal of the GMC, are shown in other bold colours. The locus of the gas which is situated in the chosen GMC at 250 Myr (middle panel) is indicated by the white dashed lines in the other panels. The total amount of gas in `clouds' is indicated on the top left of each panel, whilst the rest of the gas is more diffuse.}
\label{fig:3}       
\end{figure}

We examined the formation and destruction of GMCs in more detail in Dobbs \& Pringle 2013. The simulation in Dobbs \& Pringle includes an $m=2$ spiral component, self gravity, heating and cooling and stellar feedback. Thus there are a number of processes which contribute to the formation and dispersal of clouds. The particle mass of the simulation was 312.5 M$_{\odot}$, thus massive, $10^5$ or $10^6$ GMCs are well resolved but clouds of $\lesssim$ few $10^4$ K are not well resolved. Figure~\ref{fig:3} shows the evolution of a cloud over a period of 40 Myr. The cloud was selected at a time of 250 Myr, as shown in the middle panel, and has a mass of $2 \times 10^6$ M$_{\odot}$. The earlier timeframes (earlier panels) show that the gas which goes into the cloud is a mixture of diffuse gas and smaller clouds. Some clouds seem to simply adjoin onto the massive GMC, then move away again as the GMC disperses, without really mixing with the gas in the GMC. Similarly the GMC then disperses into smaller clouds and diffuse gas. The GMC disperses through a combination of stellar feedback and shear (Dobbs \& Pringle 2013). We also note that whilst the $2\times10^6$ M$_{\odot}$ cloud forms through the accumulation of other gas and clouds, and similarly disperses, some clouds evolve quite differently. Some clouds end their lives by being added on to the forming GMC. Conversely, some clouds are formed as a result of the GMC dispersing. Clouds tend to evolve and form stars on roughly a crossing time (Dobbs \& Pringle 2013), as supposed by Elmegreen 2000.

In forthcoming work (Dobbs et al., in prep) we investigate whether collisions are full mergers, grazing collisions, and how much gas is transferred into a resultant cloud upon a collision. We find that a substantial fraction of cloud-cloud collisions are more like grazing collisions than mergers.

\section{Star formation in grand design and flocculent galaxies}
So far we have considered galaxies with a fixed spiral potential. It is not clear how many galaxies exhibit grand design structure and how many flocculent, whilst some galaxies appear different in IR compared to optical tracers (Block \& Wainscoat 1991, Block et al. 1994). We consider in our models an extreme case where we impose no spiral potential, and the gas is subject to a completely smooth stellar potential. This produces a very flocculent spiral structure. In this case, with no spiral arms to concentrate clouds together, there are fewer cloud-cloud interactions, clouds tend to be smaller (for the same level of stellar feedback) compared to the case with spiral arms, and form largely due to gravitational instabilities (Dobbs et al. 2011).

\begin{figure}[t]
\includegraphics[scale=.65]{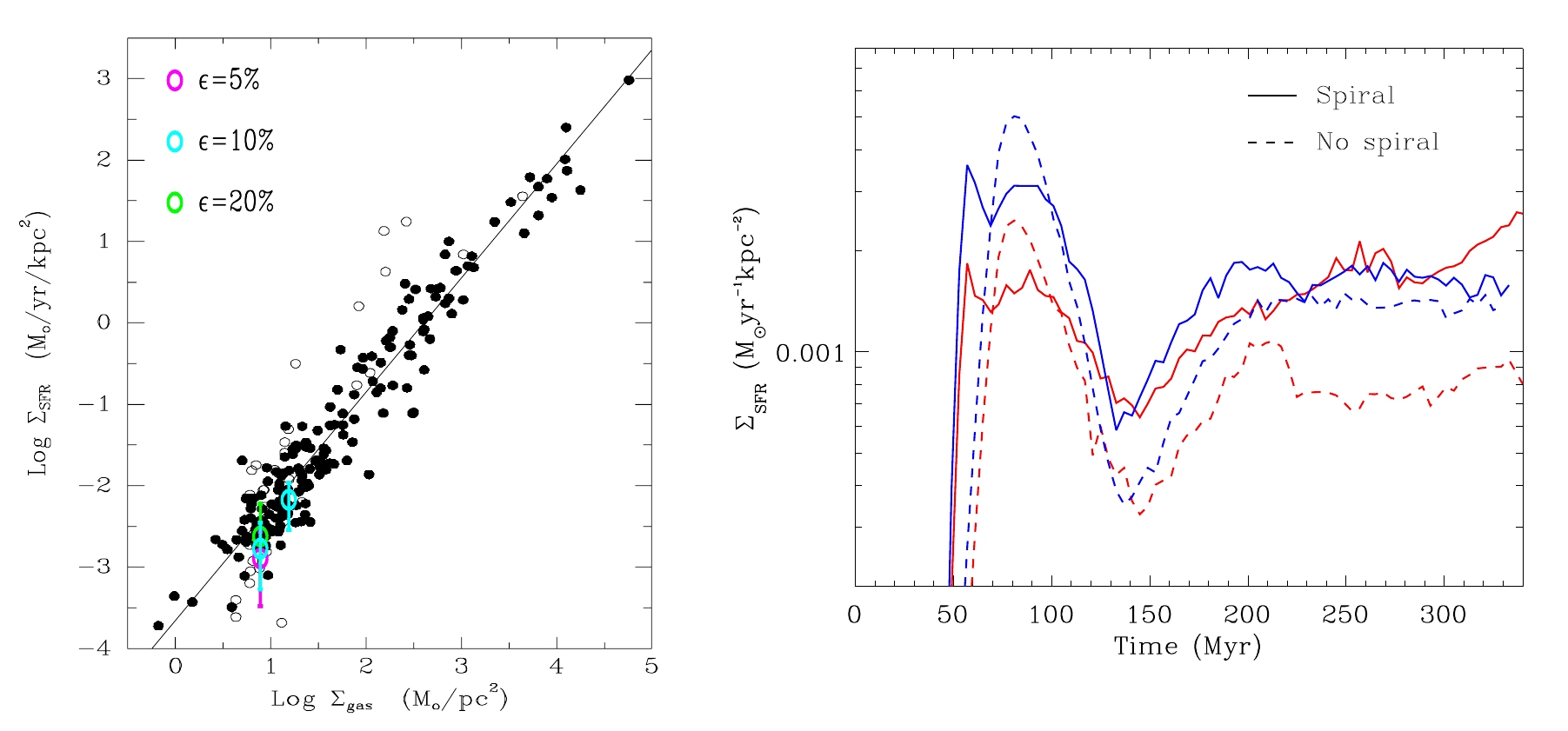}
%
%
\caption{The star formation rate is shown for simulations with different feedback efficiencies (left) and with and without a spiral potential (right). The red lines are for $\epsilon=0.05$ and the blue $\epsilon=0.1$. The black dots on the left panel are observations from Kennicutt 2008.}
\label{fig:4}       
\end{figure}

Previous observations have found that the star formation rate does not appear to significantly differ between flocculent and grand design galaxies  (Elmegreen et al. 1986; Stark et al. 1987). We have investigated the amount of star formation by comparing galaxies with different strength spiral potentials (Dobbs \& Pringle 2009) and with and without a spiral potential (Dobbs et al. 2011). Although they did not include star formation explicitly, Dobbs \& Pringle 2009 estimate the star formation rate from the mass of bound regions (and their corresponding free fall times) in the simulations. They found that the velocity dispersion increases with higher strength shocks, so the amount of bound gas (and therefore star formation) is not found to increase significantly. Dobbs et al. 2011 found that although spiral shocks have a noticeable impact on the velocity dispersion, when present stellar feedback has the greatest impact on the velocity dispersion of gas in the disc, and generally the evolution of the ISM (fractions of gas in different ISM components, density PDFs). Consequently the stellar feedback has biggest effect on the star formation rate. If  star formation is increased, more energy is injected into the ISM through stellar feedback, so that molecular clouds disperse and there is more hot gas. This in turn reduces star formation. Conversely if star formation is reduced, there is nothing to stop dense clouds gravitationally collapsing (in the absence of magnetic fields) and the star formation rate increases. In this sense, star formation is self regulating. Figure~\ref{fig:4} shows the star formation rates for simulations with different star formation efficiencies. Although there is a factor of 4 difference between the maximum and minimum efficiency, the change in star formation rate is more like a factor of 2.

In Dobbs et al. 2011, we also compared star formation rates in galaxies with and without a spiral potential. Again the spiral arms did not seem to have a large effect on the star formation rate (see Figure~\ref{fig:4}), only increasing the star formation rate by less than a factor of 2. Again, the star formation seems to be regulated by feedback, and resembles the scenario put forward (e.g. by Elmegreen et al. 1986 and Vogel et al. 1988) that spiral arms merely gather clouds together in the spiral arms. However it is evident from Dobbs et al. 2011 that the spiral arms do have an effect, namely that in gathering smaller clouds they lead to a population of more massive clouds, absent in the simulations without spiral arms. These more massive GMCs ($>10^6$ M$_{\odot}$) live longer and tend to produce a higher fraction of stars compared to their smaller mass counterparts. In Dobbs et al. 2011 this only led to a small difference in star formation rate because the number of such massive clouds was small, but potentially (e.g. with a stronger shock) the spiral arms could have more influence.

\section{Is the Milky Way flocculent or grand design?}
We touched on the spiral structure of galaxies, and the differences (or lack of differences) for star formation in the previous section. In recent work we have been investigating whether our own Galaxy, the Milky Way, is likely to be flocculent or grand design. There is still debate about the number of arms in the Galaxy (see e.g. Vall\'ee 2005) and whether the dynamics fit slowly or dynamically evolving spiral arms (Baba et al. 2011). We have recently been using the radiative code TORUS to generate synthetic HI and CO maps of simulated galaxies for comparison with observations (Acreman et al. 2010, 2012, Duarte-Cabral et al. in prep.). One application of this work has been to perform a large number of simulations (which do not include self gravity or stellar feedback and are therefore require relatively less computational time) and see which best fits ISM maps of galaxy. Pettitt et al. 2014 perform simulations specifically of spirals with fixed potentials, with a bar, 2 or 4 spiral arms and across a large parameter space of pitch angles and pattern speeds. They then carry out fitting between synthetic CO maps and the map of Dame et al. 2001. Whilst observations of ISM tracers can be used to interpret the structure of the Galaxy, using simulations has the significant advantage that there are no difficulties finding the distance to features in velocity space.

\begin{figure}
\includegraphics[scale=.75]{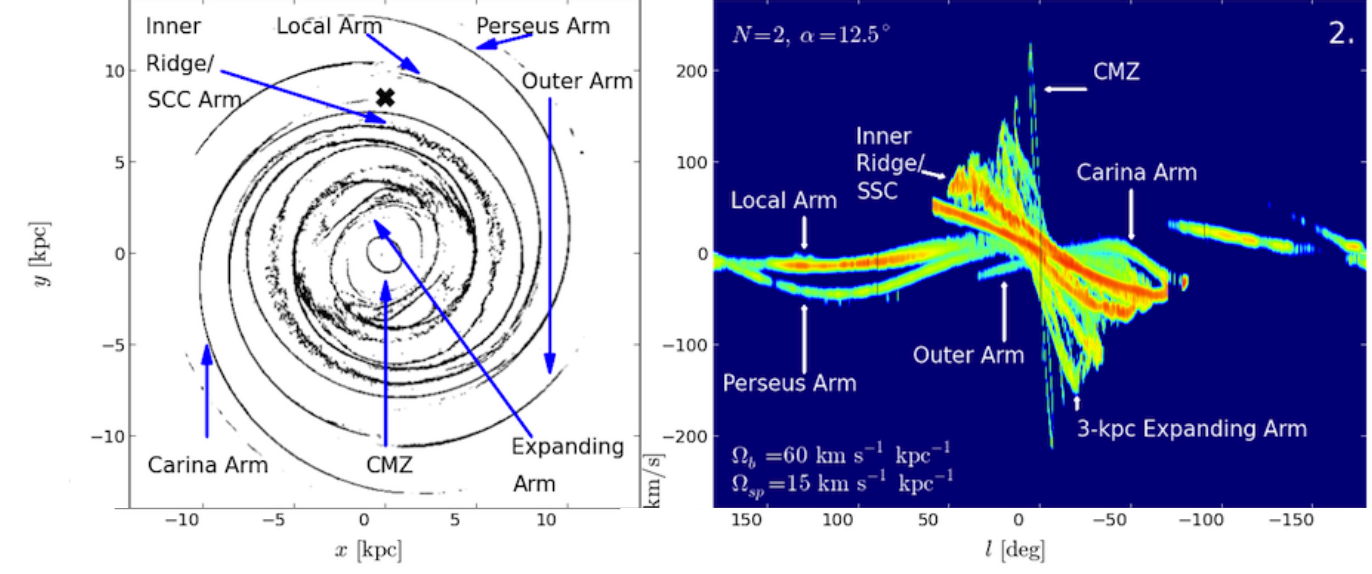}
%
%
\caption{Results are shown for a hydrodynamical model of a possible Milky Way, with a bar and 2 spiral arms with pitch angle $12.5^\circ$, and spiral arm pattern speed of 15 km/s/kpc. The map of CO emission for the model is shown on the right. The structure compared to actual observations (Dame et al. 2001) is in reasonable agreement, but the Carina arm is at the wrong velocities. We were unable to find a model which reproduced the full structure of the Milky Way with fixed potentials. From Pettitt et al. 2014.}
\label{fig:5}       
\end{figure}

Figure~\ref{fig:5} shows one of the more successful simulations from Pettitt et al. 2014 in reproducing the structure of the Galaxy in CO. Generally, we found that it was difficult to reproduce the full spiral structure (including the Local, Perseus and Outer arms) with only 2 spiral arms and a bar, and on the whole we needed 4 spiral arms. The main difficulty then for the simulations with fixed spiral potentials, is that it is difficult to reproduce the Carina arm correctly. Either the Carina arm appears at too low velocities compared to that observed (as in Figure~\ref{fig:5}), or, if the Carina arm is in the correct position, the continuation of the arm in the vicinity of the Sun leads to far more emission at $v_{los} \sim 0$km/s than observed. Thus Pettitt et al. 2014, conclude that either the Carina arm exhibits a substantial kink, or the Carina arm is not visible in CO locally, but the latter seems extremely unlikely. Consequently Pettitt et al. 2014 suggest that a transient spiral arm pattern generated by transient stellar instabilities may be better able to reproduce a more irregular spiral structure, or that a local interaction with another galaxy or satellite may have produced a large kink. Further work will examine simulations of transient spiral arms.

\section{Conclusions}
We present the results of hydrodynamic galaxy scale simulations with stellar feedback, self gravity, and ISM heating and cooling.
GMCs appear to form by a combination of the coalescence, or agglomeration of smaller clouds, and gravitational instabilities. Some low level of feedback is required to prevent clouds from becoming very gravitationally dominated, and unable to disperse as they move away from the spiral arms. A low level of feedback similarly ensures that properties of GMCs are in good agreement with observations. Star formation appears to be regulated by the velocity dispersion of the ISM, which in turn is dependent on stellar feedback and / or the spiral shock. Spiral arms do not appear to influence the star formation rate significantly in our models, rather the spiral arms simply gather the gas into more massive GMCs. 

We highlighted the evolution of one GMC in particular, and argued that GMC evolution is quite complex involving the accumulation of both smaller clouds, and more diffuse gas, whilst cloud--cloud interactions are often less disruptive than inferred by the term 'collision', often taking the form of grazing, rather than head on collisions. Finally we discuss whether the Milky Way is a grand design or flocculent spiral. It is found that is difficult to reproduce the structure of the Milky Way with static potentials, suggesting a more complex model involving an interaction and / or a transient spiral pattern is required.

\begin{acknowledgement}
CLD acknowledges funding from the European Research Council for the FP7 ERC starting grant project LOCALSTAR.
\end{acknowledgement}
%

%
%
%

\end{document}